\documentclass{ws-procs9x6}
\usepackage{epsfig,latexsym}
\usepackage{graphicx}% Include figure files
\usepackage{dcolumn}% Align table columns on decimal point
\usepackage{bm}% bold math
\begin{document}
\title{Effective Lagrangians for QCD: Deconfinement and Chiral Symmetry Restoration}
\author{\'Agnes M\'{o}csy}
\address{Institut f\"ur Theoretische Physik, J.W.
Goethe-Universit\"at, Postfach 11 19 32 60054 Frankfurt am Main,
Germany \\E-mail: mocsy@th.physik.uni-frankfurt.de}
\author{Francesco Sannino$^{\ast}$}
\address{{\rm NORDITA \& The Niels Bohr Institute},
Blegdamsvej 17
2100 Copenhagen~\/O,~Denmark\\$^{\ast}$ Speaker at the conference.\\
E-mail: francesco.sannino@nbi.dk}
\author{Kimmo Tuominen}
\address{University of Jyv\"askyl\"a, Department of Physics,  P.O. Box 35,\\
40014 Jyv\"askyl\"an Yliopisto, Finland \\
E-mail: kimmo.tuominen@phys.jyu.fi}
\maketitle \abstracts{Effective Lagrangians for Quantum
Chromodynamics (QCD) especially suited for understanding
deconfinement and chiral symmetry restoration at non-zero
temperature and matter density are reviewed. These effective
theories allow one to study generic properties of phase
transitions using non-order parameter fields without loosing the
information encoded in the true order parameter. {}For the pure
gauge theory we demonstrate that, near the deconfining phase
transition, the center group symmetry is naturally linked to the
conformal anomaly. Another relevant outcome is that when the
theory contains also quarks we can explain the intertwining of
chiral symmetry restoration and deconfinement for QCD with matter
fields either in the fundamental or in the adjoint representation
of the gauge group. As a test of our general approach we show that
our results are applicable also at non-zero baryon chemical
potential. We also predict new testable substructures to be
present in the phase diagram of quarks in the adjoint
representation of the gauge group. Here we provide some new
insights on the large $N$ limit of gauge theories by investigating
the hadronic world. We propose that the world of infinite $N$
should already be well described when $N=6$ for QCD with two and
three light flavors. Finally, we suggest possible future
applications of our results for heavy ions collisions.}

%%%%%%%%%%%%%%%%%%%%%%%%%%%%%%%%%%%%%%%%%%%%%%%%%%%%%%%%%%%%%%%%%%%%%%%
%
%Use this if your figures are put in a subdirectory having the same
%name as the main latex file, ie:
%
%      ws-procs9x6/procs-fig1.eps
%      ws-procs9x6/procs-fig2.eps
%      ws-procs9x6/procs-fig3.eps
%      ws-procs9x6/procs-fig4.eps
%      etc.
%
%\begin{figure}[htbp] %ORIGINAL SIZE: width=1.4TRUEIN; height=1.5TRUEIN
%\figurebox{}{}{procf1} %100 percent
%\caption{Labeled tree {\it T}.}
%\end{figure}
%

%%%%%%%%%%%%%%%%%%%%%%
\section{Introduction}

Symmetries, anomalous and exact, are used to constrain effective
Lagrangian theories \cite{}. The latter are applicable to any non
perturbative region of the QCD or QCD-like phase diagram, whenever
the relevant degrees of freedom and the associated symmetries are
known. To decide in which phase a strongly interacting theory can
be, one uses experimental inputs, model computations such as
Nambu-Jona Lasinio, and/or computer simulations. Exact non
perturbative constraints, such as anomaly matching conditions, are
another elegant and powerful way to help deciding among the phases
that a strongly interacting theory (vector- and chiral-like) can
be in. The original idea of t'Hooft has been extended
\cite{Sannino:2000kg,{Hsu:2000by},Sannino:2003pq} to strongly
interacting gauge theories at non-zero chemical potential. At non
zero temperature anomaly matching conditions are also expected to
be applicable. However, here corrections arise to the Ward
anomalous identity \footnote{The Adler-Bardeen theorem still
\cite{Adler:er} applies but the Sutherland-Veltman one
\cite{Sutherland:1967vf} can be violated when Lorentz symmetry
breaks \cite{Pisarski:1997bq}. The modification of the anomalous
Ward identity at non-zero temperature is due to a non-local
operator induced by temperature corrections computed using the
pion Lagrangian \cite{Pisarski:1997bq}. Why this is not a problem
at non-zero chemical potential? Take again the pion effective
Lagrangian whose coefficients now depend on the baryon chemical
potential and image to compute higher order corrections. These
corrections are similar to the one in vacuum (actually more
suppressed at large chemical potential) no violation of locality
is found with no consequence on the Sutherland-Veltamn theorem.}.
't Hooft anomaly conditions are not sufficient, in general, to
select a unique phase. A possible guide at zero chemical potential
has been suggested in \cite{Appelquist:2000qg}. Simply, one
chooses the phase, respecting anomaly matching conditions, with
the lowest number of degrees of freedom counted according to the
coefficient of the entropy. This guide works well for known
theories, such as QCD with two flavors and three colors, while it
allows a number of predictions for strongly interacting gauge
theories, such as chiral gauge theories. It would be interesting
to investigate a possible generalization to non-zero chemical
potential.

Already, a number of novel effective Lagrangians have been written
in literature to describe QCD and similar theories at zero
temperature (see \cite{Sannino:2003ff} for a review). Let us now
mention a few before turning to the problem of deconfinement and
chiral symmetry breaking.

At zero temperature and quark chemical potential a number of
quantitative non-perturbative predictions about the spectrum and
the vacuum properties of QCD with one Dirac flavor have been made
by constructing an effective Lagrangian able to interpolate from
super Yang-Mills to QCD \cite{Sannino:2003xe}. These predictions
can be tested via standard lattice simulations. Without entering
in details we can say that these results are linked to a different
type of $1/N$ expansion around the supersymmetric limit
\cite{Armoni:2003fb}, in which the fermions transform according to
the two index antisymmetric representation of the gauge group.

This expansion in the inverse of number of colors may very well be
more convergent then the ordinary $1/N$ expansion. Here the
fermions remain in the fundamental representation of the gauge
group, while increasing the number of colors. In
\cite{Harada:2003em}, for example, the existence of a critical
number of colors has been identified. For and above this number of
colors the low energy $\pi-\pi$ scattering amplitude, computed
from the sum of the current algebra and vector meson terms, is
crossing symmetric and unitary at leading order in a $1/N$
expansion. This critical number of colors turns out to be $N=6~$,
and is insensitive to the explicit breaking of chiral symmetry.
This means that ordinary $1/N$ corrections for the real world are
large. These results are supported by the findings in
\cite{Uehara:2004es}. As an important outcome, our results are
consistent with the expectation that the low lying sigma state,
$\sigma(560)~$, is not a $q\bar{q}$ object
\cite{Jaffe:1976ig,{Black:1998wt},Pelaez:2003dy,{Uehara:2003ax}}.
This has implications on the physics of chiral symmetry
restoration, since this state should not be considered as the
chiral partner of the pion, as it is at times assumed in
literature. While we expect large $1/N$ corrections for $N=3~$,
for six or more colors and with two or three flavors the physics
should be well described by the large number of colors limit, for
which a number of properties can be diagrammatically deduced
\cite{Witten:1979kh}. $SU(6)$ gauge theories are currently
explored by lattice simulations \cite{Lucini:2003nd}. Encouraged
by the findings in the hadronic world
\cite{Harada:2003em,{Uehara:2004es},{Pelaez:2003dy}}, we predict a
transition from the world of infinite $N$ to the world of small
$N$ already for $N=6$ with two flavors, and possibly with three
flavors as well. Some ideas on $N={\rm infinity}$ at non-zero
temperature are also available \cite{Pisarski:1997yh}.

Deconfinement and chiral symmetry restoration as function of
temperature, quark chemical potential or number of flavors has
always attracted much interest \cite{Brown:dm}. Here we introduce
recently developed effective Lagrangians for Quantum
Chromodynamics (QCD), relevant for gaining insight on the
deconfinement and/or chiral symmetry restoration problem at
non-zero temperature and matter density. In literature one can
already find a large number of models which try to understand/fit
lattice results. The goal we have here is to provide a unifying
point of view in which different models can be seen as different
description of the same physics. At the same time we try to
develop new ideas and tools to investigate phase transitions.
Clearly, some of the presented tools need to be sharpened in the
future.

We first consider the effective theory unifying two apparently
very different sectors of a generic Yang-Mills theory at non-zero
temperature \cite{Sannino:2002wb}: the hadronic sector and the
Polyakov loop. This theory is able to communicate the information
about the center group symmetry to the hadronic states. It also
provides the link between deconfinement and conformal anomaly. The
latter has already been used for years in literature for
describing the deconfinement phase transition in pure glue.

The basic general idea is to generalize the Landau theory by
including non-order parameter fields. Technically, we {\it
integrate in} a heavy field. We will see that this allows us to
study generic properties of phase transitions using the non-order
parameters field. Effective Lagrangians for strongly interacting
theories with matter fields will also be reviewed here. Via these
theories we offer a simple, economical and unifying interpretation
of the intertwining between chiral symmetry restoration and
deconfinement in QCD with matter fields either in the fundamental,
or in the adjoint representation of the gauge group, as function
of temperature and/or chemical potential. We also show that the
most relevant term is a trilinear interaction between the singlet
field and the order parameter. This allowed interaction term
differentiates between the different fermion representations, and
it has been neglected previously in the literature
\cite{Dumitru:2000in,{Rischke:2003mt}}. We finally suggest
possible physical applications.

%%%%%%%%%%%%%%%%%%%%%%%%%%%%%%%%%%%%%%%%%%%%
\section{Heavy Fields and Phase Transitions}
The phase transition dynamics and the associated critical behavior
are best investigated using order parameters. These are the
degrees of freedom whose correlation length diverges when
approaching the phase transition. However, the choice of the order
parameters is not always obvious. Furthermore, it often turns out
that even if the order parameter can be formally constructed, this
may be hard, or even impossible to be directly detected
experimentally. A time honored example is the Polyakov loop in
gauge theories. The characteristic feature of an order parameter
is that is zero in the symmetric phase, but attains a finite
non-zero value in the symmetry broken phase. In what follows we
will term order parameter field a field whose expectation value is
a true order parameter, that has the characteristic behavior
described above. Any field whose expectation value does not have
this kind of behavior will be called a non-order parameter field.
{}For simplicity, in the following we will consider non-order
parameter fields which are singlets under the symmetry
transformations acting on the order parameter.

One can ask some simple questions: Given a system consisting of an
order parameter and singlet field(s), what can we learn about the
phase transition by monitoring the singlet field?  Can we identify
the onset of the phase transition without referring to the order
parameter? We will show that there is a clear and universal
characteristic behavior of the singlet field, induced by the order
parameter close to the phase transition. Our considerations are
universal, and as such, can be carried over to virtually any phase
transition once the symmetries of the order parameter are
identified. We will explicitly consider the cases of $Z_2$
symmetry, which is of relevance for the pure Yang--Mills gauge
theory with two colors, and $SU(4)$ which is of relevance for the
effective theories of two color QCD with two quark flavors in
either the fundamental or adjoint representation of the gauge
group.

Lattice simulations of the pure Yang-Mills gauge theory already
confirm our prediction, that the behavior of the order parameter
close to the critical temperature is reflected in the behavior of
the non-order parameter field.

For QCD with quarks lattice simulations pose the following
interesting puzzle: Why, for matter in the fundamental
representation deconfinement and chiral symmetry restoration
appear to be linked with a single phase transition observed at a
given critical temperature, while for matter in the adjoint
representation there are two phase transitions, well separated in
temperature? We will show how our effective Lagrangian description
can offer a simple unifying way of addressing this puzzle.

%%%%%%%%%%%%%%%%%%%%%%%%%%%%%%%%%%%%%%%%%%%%%%%%%
\section{From the Polyakov Loop to the Glueballs}
%%%%%%%%%%%%%%%%%%%%%%%%%%%%%%%%%%%%%%%%%%%%%%%%%%%%%%%%%%%%%%%%%%%%%%%%%%
\subsection{A Unifying Model: Center Group Symmetry and Conformal Anomaly}
Consider the pure Yang--Mills theory with $SU(N)$ gauge symmetry.
At non-zero temperature the Yang-Mills theory possesses a global
$Z_N$ symmetry. This symmetry is intact at low temperatures and is
broken at high temperatures. The associated order parameter is the
Polyakov loop, which is the trace of the thermal Wilson line.
Under the action of $Z_N$ the Polyakov loop transforms as
$\ell\rightarrow z\ell$ with $z\in Z_N$.

The Polyakov loop condensation is associated to deconfinement. The
reason being that the potential $V(\vec{x},T)$ at a given
temperature $T$ between static fundamental charges is related to
the Polyakov loop:
\begin{eqnarray}\nonumber
  \label{eq:fpotential}
\exp(-V(\vec{x},T)/T) \equiv \langle \ell(\vec{0}) \ell^{\dagger}
(\vec{x})\rangle \quad_{\longrightarrow \atop
{|\vec{x}|\rightarrow \infty}} \quad |\ell|^2 \quad ,
\end{eqnarray}
{}For $T<T_c$ the static quarks experience a linearly rising
potential
\begin{eqnarray}
V(\vec{x},T)=\sigma(T)\,|x| \ ,
\end{eqnarray}
where $\sigma(T)$ is the string tension. {}For $T>T_c$ the
potential is no longer a confining one:
\begin{eqnarray}
V(\vec{x},T) \propto k(T) \ ,
\end{eqnarray}
and the potential does not depend on the distance between the
static sources, here chosen to be in the fundamental
representation of the gauge group.

Close to $T_c$, using age-old arguments by Landau, one can write a
mean field effective potential of the form
\begin{eqnarray}
V(\ell)=T^4 \mathcal{F}(\ell),
\end{eqnarray}
where $\mathcal{F}(\ell)$ is a $Z_N$-invariant polynomial in
$\ell$. It was Svetitsky and Yaffe who provided the first reliable
non-perturbative study of the critical behavior of Yang-Mills
theories as well as the form of the effective potential displayed
above \cite{Svetitsky:1982gs,Yaffe:qf,Svetitsky:1983cc}. The idea
of describing the Yang-Mills pressure using directly a mean field
theory of Polyakov loops has been recently advocated by Pisarski
\cite{Pisarski:2000eq}. This model has since been used in several
phenomenological studies \cite{Dumitru:2000in,Scavenius:2002ru},
which miss to include of a relevant term
\cite{Mocsy:2003tr,Mocsy:2003un,Mocsy:2003qw}. The scale dimension
is set by the temperature. The physical states (i.e. the hadrons),
however, do not carry any charge under $Z_N$. So in order for a
hadronic state to know about the center group symmetry this will
have to communicate with the Polyakov loop. Besides, approaching
the critical temperature from the confining side, the relevant
degrees of freedom are the hadronic states which for the pure
Yang--Mills theory are the glueballs.

In the following, we consider as important representative of the
hadronic spectrum the lightest scalar glueball. At zero
temperature the well known effective theory constrained by the
trace anomaly has been constructed in \cite{Schechter:ak}, using
the potential of the form
\begin{eqnarray}
H\ln \frac{H}{\Lambda^4} \ .
\end{eqnarray}
The glueball field $H$ is related to $G_{\mu\nu}^a\,G^{\mu\nu,a}$,
where $G^{\mu\nu,a}$ is the gluon field strength and $a=1,\ldots,
N^2-1$ the gauge indices. It has been shown that this potential
encodes the basic properties of the Yang--Mills vacuum at $T=0$
\cite{Sannino:1997dd}, and it has also been used at non-zero
density \cite{Brown:2001nh} and temperature
\cite{{Agasian:fn},Carter:1998ti}. As we increase the temperature
the Polyakov loop becomes a well defined object. We stress that
sofar in the literature the Yang-Mills pressure has been described
by either glueball theories \cite{Agasian:fn} and their
generalizations \cite{Agasian:2003ux}, or directly with the
Polyakov loop model \cite{Dumitru:2000in}. {}In effective glueball
theories a possible drop at $T_c$ of the {\it non-perturbative}
contribution to the gluon condensate has been often considered as
an indication of deconfinement \cite{Agasian:fn,{Brown:2001nh}}.
To be more precise, since temperature breaks Lorentz invariance,
one should consider independently the non-perturbative
contribution to the trace anomaly of $G^{0i}G_{0i}$ and
$G^{ij}G_{ij}$. In the vacuum we know that the expectation value
of the second component dominates since confinement is associated
to condensation of magnetic type charges. It is this component
which should decrease at the phase transition. However, since the
gluon condensate is not an order parameter for the center group
symmetry there is, a priori, no guarantee that such a drop, even
if observed, should appear at the same critical temperature.
Another important point is that the glueball theory by
construction is blind to the number of colors. This explains why
it always predicts a first order phase transition independent of
the number of colors of the underlying Yang-Mills theory  one
wants to describe \cite{Scavenius:2002ru}.

The Polyakov model, on the other hand, has automatically built in
the knowledge about the number of colors, but looses contact with
a simple physical picture in terms of hadronic states.

We will now show, by marrying these two theories, that: i) The
drop of the non-perturbative part of the gluon condensate {\it
must} happen exactly at the deconfining phase transition
\cite{Sannino:2002wb}, partially justifying the glueball models;
ii) The information about the order of the transition is now
encoded in a nontrivial way in the profile of the condensate drop.

\begin{figure}[h]
\begin{center}
\begin{minipage}[t]{6cm}
\epsfig{file=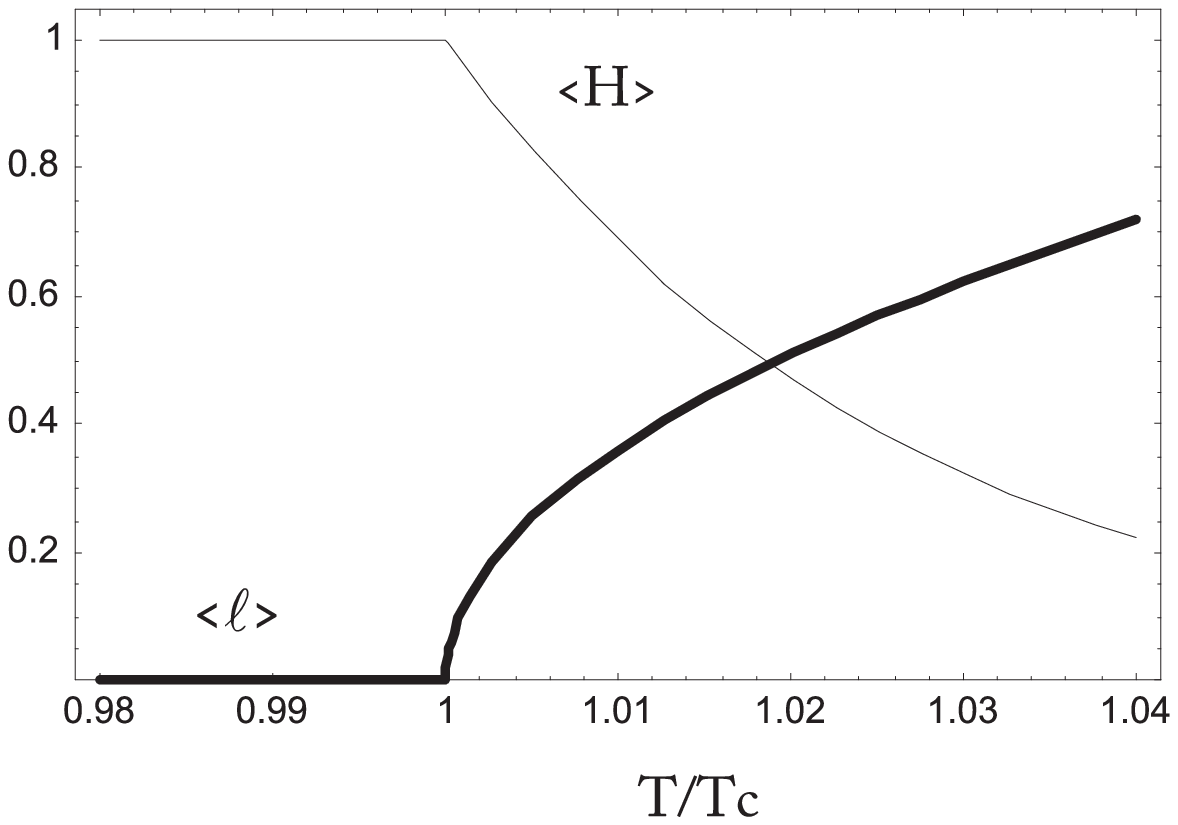,height=4cm,width=5cm}
\end{minipage}%
\begin{minipage}[t]{6cm}
\epsfig{file=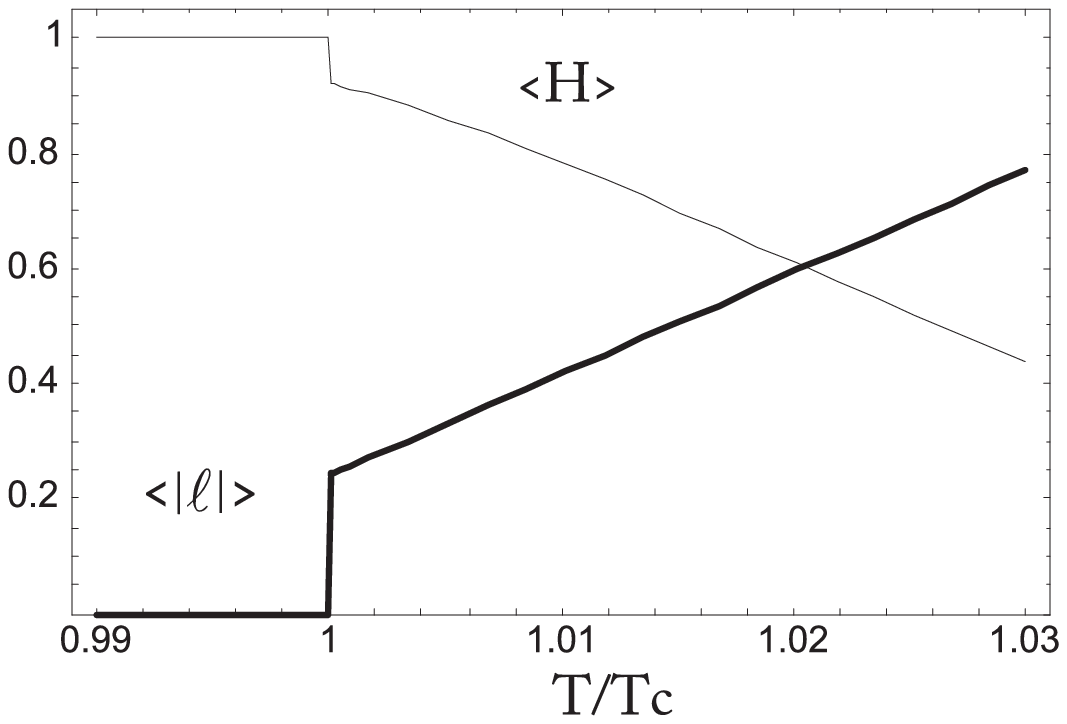,height=4cm, width=5cm}
\end{minipage}
\end{center}
\caption{The behavior of the expectation values of the fields as a
function of temperature. The left panel shows the two color case
with second order transition. The right panel shows the three
color case with the first order transition, characterized by a
finite discontinuity in the expectation values.} \label{tree}
\end{figure}

We start with the following general potential constructed using
all of the relevant symmetries, i.e. trace anomaly, $Z_N$
invariance and analyticity of the interaction term between the
glueball field and the Polyakov loop\cite{Sannino:2002wb}
\begin{eqnarray}
V(\ell,H)=H\ln\frac{H}{\Lambda^4}+V_T(H)+H\mathcal{P}(\ell) +
T^4\mathcal{F}(\ell)\, .
\end{eqnarray}
Here $V_T(H)$ is the intrinsic temperature dependence of the
glueball gas, which we can neglect in what follows, since the
glueball is heavy near the phase transition. The most general
interaction term compatible with the saturation of the trace
anomaly is $H\mathcal{P}(\ell)$, where $\mathcal{P}(\ell)$ is a
$Z_N$ symmetric polynomial in $\ell$. In the static limit we can
use the equation of motion for the $H$ field to integrate this out
in terms of $\ell$ field, yielding the functional relation
$H\equiv H(\ell)$.  This shows that the center group symmetry is
transferred to the singlet sector of the theory. However, we find
it more illuminating to keep both fields. {}For two colors we take
\begin{eqnarray}
\mathcal{F}(\ell) &=& a_1\ell^2 + a_2\ell^4+\mathcal{O}(\ell^6) \
, \nonumber \\ \mathcal{P}(\ell) &=& b_1\ell^2 +
\mathcal{O}(\ell^4)  \ ,\nonumber
\end{eqnarray}
and the coefficients $a_2$ and $b_1$ are assumed to be positive
and independent of the temperature, while the coefficient $a_1$ is
taken to be $a_1(T)=\alpha (T-T_*)$ with $\alpha > 0$. Using these
expressions in the potential and solving for the minimum, one
finds that below $T_c$ the two fields are completely decoupled and
\begin{eqnarray}
\langle \ell\rangle = 0, \quad \langle H\rangle=\Lambda^4/e\, .
\end{eqnarray}
As the critical temperature
$T_c=T_*+\frac{b_1}{e\alpha}\frac{\Lambda^4}{T_c^3}$ is reached,
the symmetry is spontaneously broken and the Polyakov loop obtains
a non-zero expectation value, which affects also the one for $H$:
\begin{eqnarray}
\langle \ell\rangle\propto \frac{\Delta T}{T_c}, \quad \langle
H\rangle=\frac{\Lambda^4}{e}\, \exp\left[-2 b_1 \langle \ell
\rangle ^2 \right]\, .
\end{eqnarray}
For three colors the essential modification is the inclusion of
the $\ell^3$ and $\ell^{* 3}$ terms into the Polyakov loop
potential $\mathcal{F}~$, while the interaction potential becomes
$\mathcal{P}\propto |\ell|^2$. The cubic terms render the
transition first order, and thus the change in both the Polyakov
loop and the glueball becomes discontinuous. Once more we deduce
that the change in the order parameter $\ell$ induces a change in
the expectation value of the non-order parameter field. The
behavior in these two cases is shown in figure \ref{tree}.
We then conclude that, not only the drop of the non-perturbative
part of the gluon condensate knows about the Yang-Mills phase
transition, but also that the drop occurs at $T_c$. Furthermore,
the profile of the gluon condensate as function of temperature
encodes the information on the order of the phase transition.

This theory unifies two apparently very different pictures of the
deconfining phase transition.

%%%%%%%%%%%%%%%%%%%%%%%%%%%%%%%%%%%%%%%%%%%%%%%%%%%%%%%%%%%%%%%
\subsection{Beyond the Static Limit: A Renormalizable Approach}
Now we will improve on some of the approximations made in the
previous section. We consider fluctuating fields and hence we
include the kinetic terms. Furthermore, in order to keep the
discussion as general as possible we will not confine ourselves to
the Yang-Mills theory. We start with a renormalizable Lagrangian
for two real scalar fields constrained only by the symmetries and
renormalizability. We consider the case of an exact $Z_2$
symmetry. The generalization to a $Z_N$ or an $O(N)$ symmetry is
straightforward. The order parameter field is $\chi$ which
transforms according to $\chi \rightarrow z\chi$ with $ z \in
Z_2$. We will consider two cases: A time independent,
$\chi=\chi(x)$, and a time dependent one, $\chi=\chi(t,x)$. The
former being relevant for deconfinement in QCD when identified
with $\ell~$, while the latter can be of use if identified with
the electroweak Higgs field for example. The most general
potential, in both cases restricted by symmetry and
renormalizability, is\footnote{There is also a linear term in $h$
whose effects up to one loop have been studied in detail in
\cite{Mocsy:2003un} and are shown not to affect our results.}:
\begin{eqnarray} V(h,\chi) =
\frac{m_h^2}{2} h^2 + \frac{m^2_{\chi}}{2}\chi^2 +
\frac{\lambda}{4!}\chi^4 + \frac{g_{1}}{2}h\chi^2 +
\frac{g_2}{4}h^2\chi^2 + \frac{g_3}{3!}h^3 +
\frac{g_4}{4!}h^4\nonumber
\end{eqnarray}
This potential can be considered as a truncated version of the
full glueball theory of the previous section, after having defined
$H=\langle H\rangle (1+h/(\sqrt c\langle H\rangle))$ and
$\chi=\sqrt\kappa\ell$. Consider first the case of
time-independent order parameter, $\chi=\chi(x)$. In this case
only the zero mode of $h$ is relevant and the Lagrangian for the
three dimensional theory is
\begin{eqnarray}
-{\mathcal L}_{3D}&=&\frac{1}{2}\nabla
h\nabla h + \frac{1}{2}\nabla\chi\nabla\chi + \frac{1}{2}m_h^2h^2
+ \frac{1}{2}m^2_\chi\chi^2\ \nonumber \\ &+&
T\frac{\lambda}{4!}(\chi^2)^2 + \sqrt T
\frac{{g_{1}}}{2}h\chi^2 +
T\frac{g_2}{4}\,h^2\chi^2 + \sqrt T\frac{g_3}{3!}h^3 +
T\frac{g_4}{4!}h^4.
\end{eqnarray}
We are interested in evaluating the corrections to the mass of the
singlet field $h$ induced by the fluctuations of the order
parameter. These are dominated at one loop by the diagram
\begin{eqnarray}
\parbox{15mm}{\includegraphics[width=15mm,clip=true]{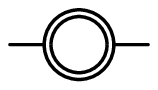}}&=&
T\,(\frac{g_1}{2})^2\int\frac{d^3k}{(2\pi)^3}\frac{1}{(k^2+m_\chi^2)^2}
=  T\, \frac{g^2_1}{32\pi m_{\chi}} \, . \label{one-loop}
\end{eqnarray}
Due to the infrared singularity this one-loop result breaks down
at $T_c~$. On general grounds though, discussed in detail in
\cite{Mocsy:2003un}, we expect the mass of $h$ to stay finite.
Therefore, further improvements are required. In order to
determine the fixed points and the associated critical exponents
we can use the $4-\epsilon$ analysis. This is currently under
investigation \cite{Mocsy}. Here, however, we shall present a
simple resummation scheme which is exact in the case of the $O(N)$
symmetric theory in the large $N$ limit. Nota bene, that although
we expect this simple model to be at best only qualitatively
correct, as we shall see, it reproduces the general features of
rather ``old'' data \cite{Bacilieri:1988dj} surprisingly well.

\begin{figure}[h]
\includegraphics[width=11truecm, height= 5truecm]{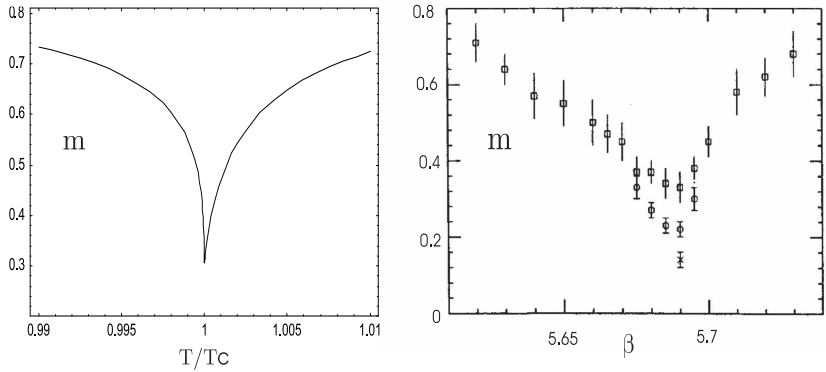}
\caption{Left panel: mass of the singlet field near $T_c$
evaluated using (\ref{sym}) and (\ref{broke}). For this plot we
chose $g_1^2/(m^2\lambda)\sim 0.9$ and $m_\chi^2/[(T_c-T)T]\sim
0.004$. Right panel: Lattice data from
\protect\cite{Bacilieri:1988dj}.} \label{massdecrease}
\end{figure}

The finite result is given by the sum of simple bubble diagrams,
yielding for the mass for the singlet field at $T_c$
\begin{eqnarray}
m_h^2(T)=m_h^2-T\frac{g_1^2}{16\pi m_\chi +\lambda T}
\buildrel{T\rightarrow T_c}\over{\longrightarrow}
m_h^2-\frac{g_1^2}{\lambda}\, . \label{sym}
\end{eqnarray}
A similar bubble sum can be performed in the broken phase with the
replacement $m_\chi^2\rightarrow 2|m_{\chi}|^2$, leading to the
same value for the mass of the singlet field at $T_c$. The
critical behavior can be traced via the slope of the screening
mass, which is simply
\begin{eqnarray}
{\mathcal{D}}^\pm \equiv \lim_{T\rightarrow T_c^\pm}
\frac{1}{\Delta m_h^2(T_c)}\frac{d\Delta m_h^2(T)}{dT}\sim
t^{\nu/2-1},
\end{eqnarray}
where $\Delta m_h^2 \equiv m_h^2(T)-m_h^2$ and the result is exact
only in the limit of large $N$. It is interesting to note that
there exists other classes of diagrams which can be exactly
resummed, and allow us to depart from the large $N$ limit. In the
broken phase such class is given by the diagrams
\begin{equation}
\includegraphics[width=6.5cm,clip=true]{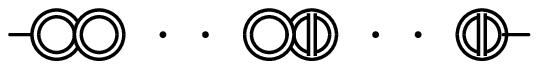}
\nonumber
\label{bubbles2}
\end{equation}
These can be resummed exactly \cite{Mocsy:2003un} yielding
\begin{eqnarray}
m_h^2(T)= m_h^2-\frac{g^2_1{\mathcal
I}}{2}\,\frac{1+\frac{\lambda}{3}{\mathcal
I}}{1+\frac{\lambda}{2}{\mathcal I}+ \frac{\lambda^2}{6}{\mathcal
I}^2}\, ,\label{broke}
\end{eqnarray}
where
\begin{eqnarray}
{\mathcal I}=\frac{T}{8\pi\,\sqrt{2}|m_{\chi}|}\, .
\end{eqnarray}
At the critical temperature, once again we find
${m_h^2(T_c)=m_h^2-\frac{g_1^2}{\lambda}}$. The qualitative
behavior of the screening mass is shown in figure
\ref{massdecrease}.
On the left panel is our schematic plot. On the right panel is
lattice data from \cite{Bacilieri:1988dj} showing the behavior of
the scalar glueball screening mass in pure SU(3) gauge theory.
Such drop in the glueball screening mass has been further
identified in \cite{Datta:1998em} for both SU(2) and SU(3).

In the case of the time-dependent order parameter field,
$\chi=\chi(t,x)$, all the modes contribute to the analysis at
finite temperature. We have shown \cite{Mocsy:2003un} that the
pole mass of the singlet field is not affected by the phase
transition, while the screening mass is still IR sensitive.
Furthermore, the screening mass of the singlet field shows the
same behavior as described above.

We stress that the presence of the trilinear term $h\chi^2~$ plays
a vital role and should not be neglected.

%%%%%%%%%%%%%%%%%%%%%%%%%%%%%%%%%%%%%%%%%%%%
\section{Chiral Symmetry versus Confinement}
\begin{figure}[h]
\begin{center}
\begin{minipage}[t]{6cm}
\epsfig{file=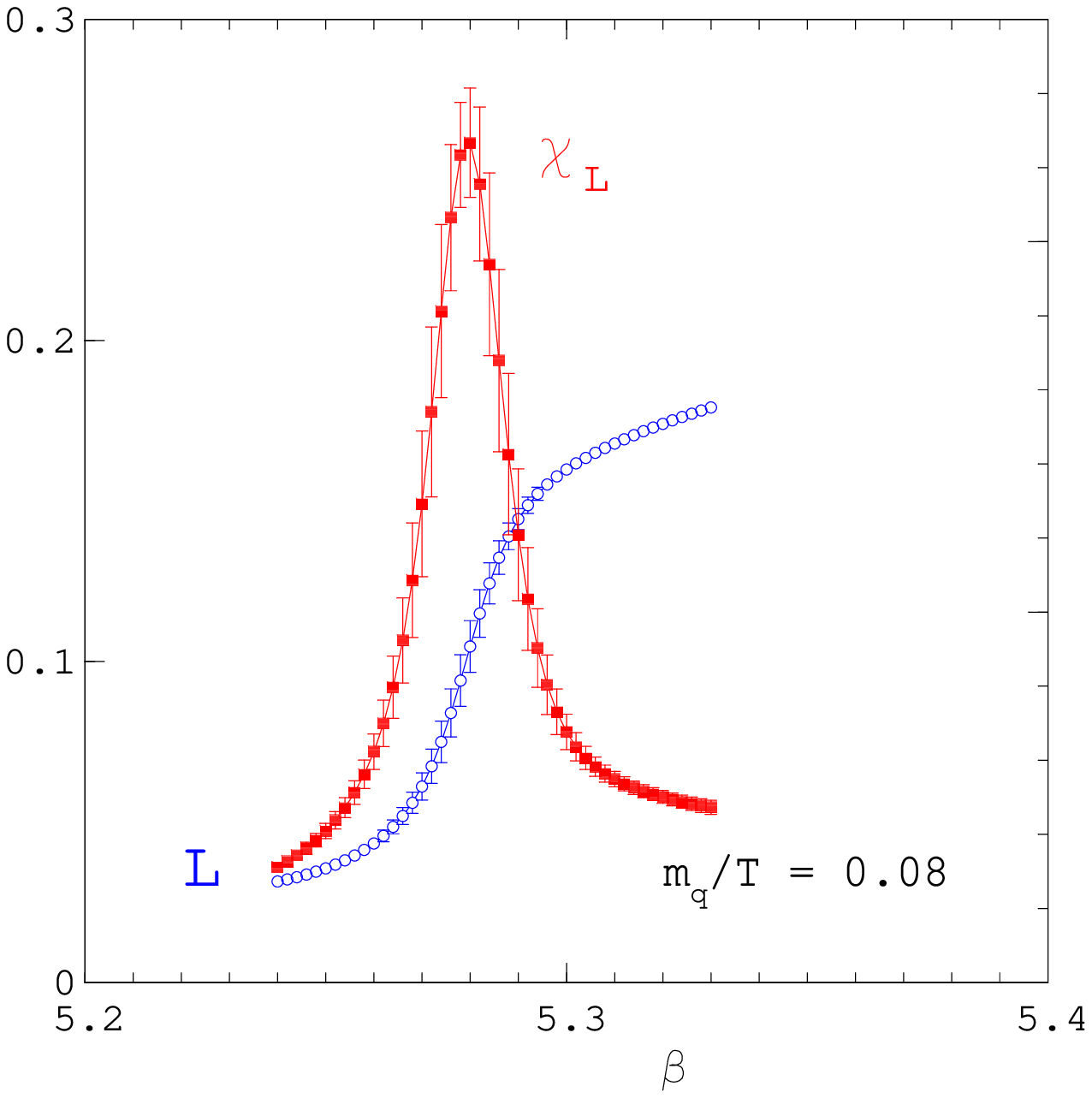,height=4cm,width=5cm}
\end{minipage}%
\begin{minipage}[t]{6cm}
\epsfig{file=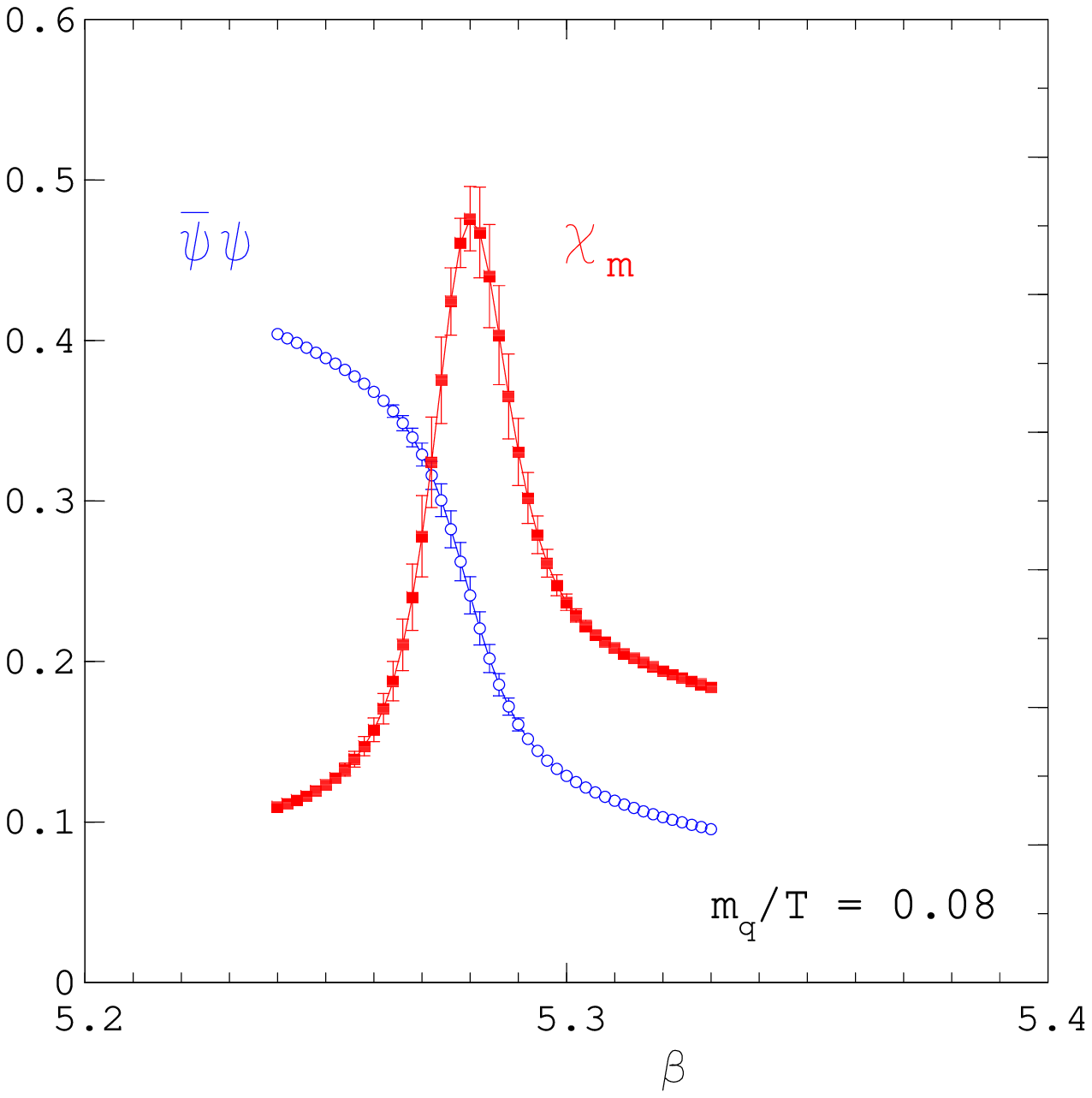,height=4cm, width=5cm}
\end{minipage}
\end{center}
\caption{Behavior of the Polyakov loop (left panel) and of the
chiral condensate (right panel) and the corresponding
susceptibilities, as determined on the lattice with quarks in the
fundamental representation. From \protect\cite{Karsch:2001cy}.}
\label{QCD}
\end{figure}

\begin{figure}[h]
\begin{center}
\begin{minipage}[t]{6cm}
\epsfig{file=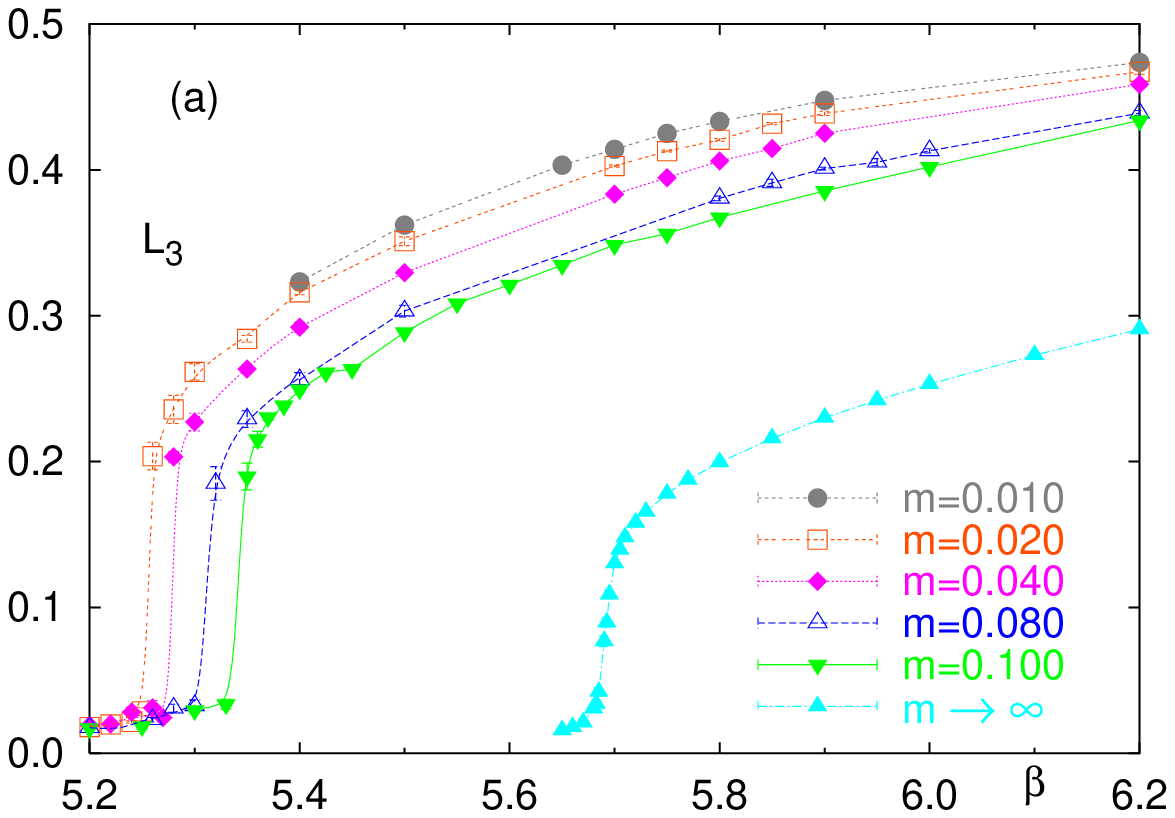,height=4cm,width=5cm}
\end{minipage}%
\begin{minipage}[t]{6cm}
\epsfig{file=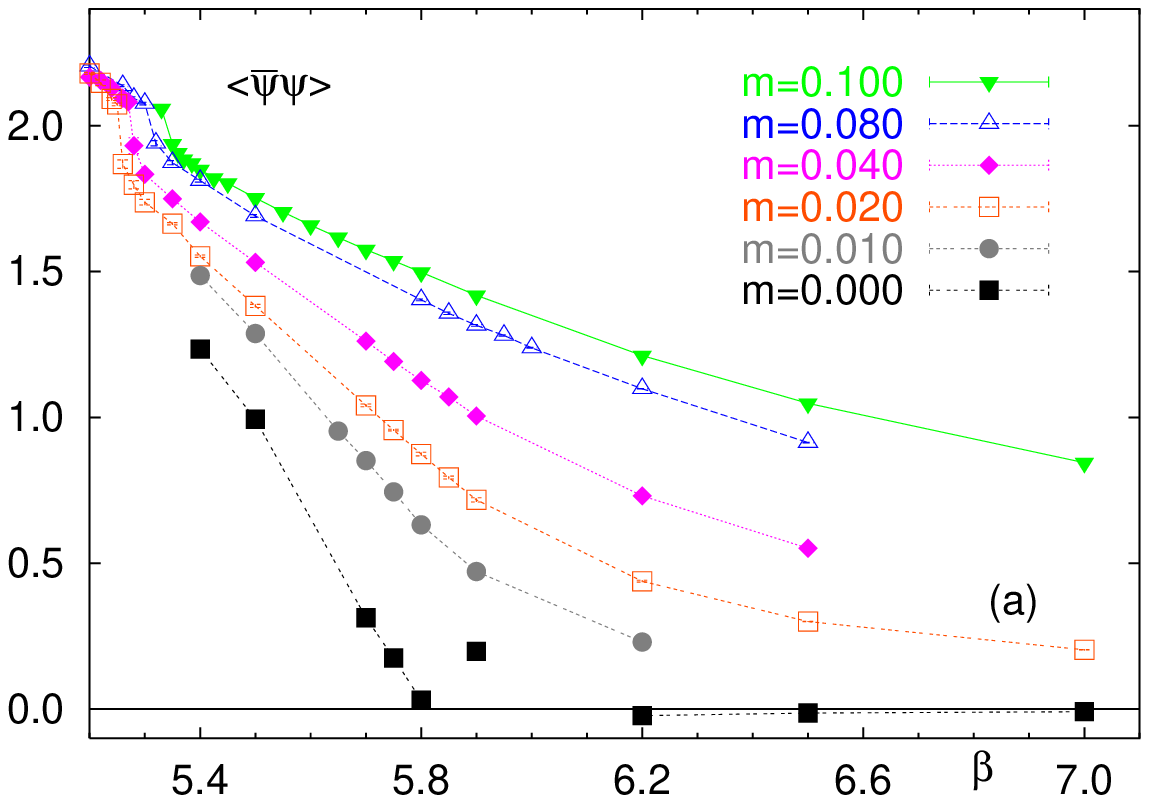,height=4cm, width=5cm}
\end{minipage}
\end{center}
\caption{Behavior of the Polyakov loop (left panel) and of the
chiral condensate (right panel) as determined on the lattice with
quarks in the adjoint representation. From
\protect\cite{Karsch:1998qj}.}\label{aQCD}
\end{figure}

Consider now adding quarks into the Yang-Mills theory. Lattice
results for quarks in the fundamental representation of the gauge
group show a clear indication that as chiral symmetry is restored,
the expectation value of the Polyakov loop rises, signalling
deconfinement. Figure \ref{QCD} from \cite{Karsch:2001cy}
illustrates that the critical temperatures\footnote{The (pseudo)
critical temperatures are identified from the peak position of the
corresponding susceptibilities, also shown in figure
\protect\ref{QCD}.} of the deconfinement and of the chiral
symmetry restoration coincide, $T_{\rm chiral} = T_{\rm deconf}~$.
We also learn from lattice results that for quarks in the adjoint
representation this is not the case, as clearly indicated in
Fig.~\ref{aQCD} taken from \cite{Karsch:1998qj}.
Here the critical temperature for chiral symmetry restoration is
$~T_{\rm chiral}\simeq 8T_{\rm deconf}$. Note also, that even if
the two transitions happen separately the chiral condensate knows
about deconfinement, as the jump in its behavior at $T_{\rm
deconf}$ seems to indicate.

Lattice simulations are already available for two color QCD at
non-zero baryon chemical potential, and observe deconfinement for
2 color QCD and 8 continuum flavors at $\mu\neq 0$
\begin{figure}
\begin{center}
\epsfig{file=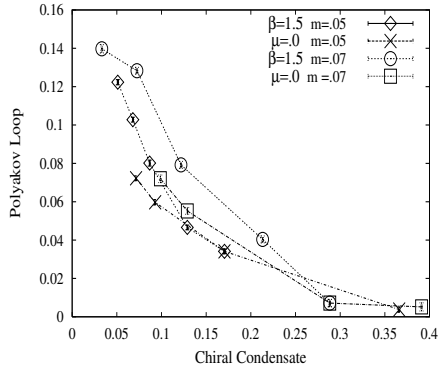,height=5cm, width=6cm} \caption{Polyakov
loop versus chiral condensate for two color QCD and 8 continuum
flavors at non-zero baryon chemical potential as determined on the
lattice. From \protect\cite{Alles:2002st}.} \label{fig:ptvspbp}
\end{center}
\end{figure}
One of the relevant points is that the Polyakov loop rises when
the chiral condensate vanishes, and the two phase transitions
happen at the same value of the chemical potential i.e.
${\mu}_{\rm chiral} = {\mu}_{d}$ \cite{Alles:2002st}.

Our goal is to provide a simple, economical and unified way to
describe all of these features. This is possible thanks to a
crucial interaction term which has been neglected in
phenomenological investigations \cite{Dumitru:2000in}, and also in
explanations of the coincidence of chiral and deconfining phase
transition \cite{Rischke:2003mt}.

%%%%%%%%%%%%%%%%%%%%%%%%%%%%%%%%%
\subsection{General Observations}
The quark representation with respect to the gauge group plays an
important role \cite{Banks:1983me}. In fact, when quarks are in
the fundamental representation of the gauge group the $Z_N$
symmetry is no longer exact for any finite value of the quark
mass. When, on the other hand, quarks are in the adjoint
representation of the gauge group the center group symmetry is
intact.

Chiral symmetry is explicitly broken by a mass term for quarks. In
the non-perturbative regime we do not know \footnote{Unless a fit
to lattice data \cite{Karsch:2000kv} is performed
\cite{Dumitru:2003cf}.} the amount of $Z_N$ breaking, which is due
to introduction of quarks in the fundamental representation.
Therefore we cannot establish which symmetry is more broken for a
given quark mass.

{} Fortunately, at least theoretically, we can take limits in the
QCD parameter space, allowing us to make exact statements
independently of the lattice results. The first limit we consider
is the one in which chiral symmetry is exact, i.e. quark masses
are zero. {}For any number of flavors and colors, $Z_N$ is not a
symmetry and the only order parameter is the chiral condensate
($\bar{q}q \sim \sigma$). We will demonstrate in the following
sections, that due to the presence of a relevant $Z_N$ symmetry
breaking term the Polyakov loop\footnote{Nota bene, that we named
the canonically normalized field associated with the Polyakov loop
$\chi$. Any quantity having a subscript $\chi$ is associated to
the Polyakov loop and not to chiral symmetry. Chiral symmetry
related quantities have the subscript $\sigma$.} has an induced
critical behavior driven by the chiral phase transition. This term
is the most relevant term differentiating among fermion
representations. Unfortunately, previously it has been omitted in
the literature \cite{Dumitru:2000in,{Scavenius:2002ru}}. We expect
that its inclusion drastically affects phenomenological
predictions and their implications for heavy ion collision
experiments.

If the $Z_N$ symmetry is only softly broken one might expect
another almost phase transition for deconfinement
\cite{Shuryak:1981fz,{Pisarski:mq},Manohar:1983md}. The main
motivation for having advocated, in the past, two independent
phase transitions is related to the possibility of two independent
scales in QCD: One associated to chiral symmetry breaking and the
other associated to deconfinement. The underlying origin of two
different scales is however unclear. Besides, as explained above,
lattice results now dismiss this possibility for QCD. A two phase
transitions situation is still possible in QCD with zero quark
masses in the limit of large number of colors for fixed number of
flavors, i.e. $N_f/N \ll 1~$, but it is unnatural in the case
$N_f\sim N$.

When turning on the quark masses the situation is less
transparent. If the quark masses are smaller than the confining
scale of the theory and barring {\it accidental} dynamical
suppression we expect that by comparing the following ratios
\begin{eqnarray} m_{\rm quark}/\Lambda \quad {\rm chiral~breaking}
\quad {\rm and} \quad N_f/N\quad {\rm Z_N~breaking} \
,\end{eqnarray} we can have a rough estimate of which symmetry is
mostly restored. It might be more accurate to use as ratio for
chiral symmetry $m_{\pi}/4\pi F_{\pi}$. Note that we expect the
$Z_N$ breaking terms to be a more complicated function of number
of flavors, colors and quark masses $f(N_f,N_c,m_{\rm
quark}/\Lambda)$. Near the chiral limit we can clearly drop the
quark mass dependence.
% Recently via a fit to
%lattice simulations in \cite{Dumitru:2003cf} it has been argued
%that the simplest $Z_3$ symmetry breaking coefficient is small and
%hence that the $Z_3$ symmetry is a better symmetry than chiral
%symmetry. Since the pion masses range between $400~MeV$ and
%$1~GeV$ this result is perhaps consistent with the fact that we
%are not at all near to the chiral limit. If however such a small
%coefficient persists even at small masses this might indicate that
%he naive counting $N_f/N$ is incorrect.

The situation for the fermions in the adjoint representation of
the gauge group will be studied below in some detail along with
the study of quarks in the fundamental representation and in the
limit of exact chiral symmetry.

We have chosen, in order to illustrate our results, to study the
two color theory with $N_f$ flavors. The rational behind this
choice is that at the same time and with a minor modification of
the effective Lagrangian we can discuss the theory with quarks in
the fundamental and adjoint representation at non-zero temperature
or quark chemical potential. The generalization to three colors is
straightforward at non-zero temperature while a bit more involved
for the quark chemical potential.

%%%%%%%%%%%%%%%%%%%%%%%%%%%%%%%%%%%%%%%
\subsection{Fundamental Representation}
The global quantum symmetry group for two flavors and two colors
is $SU(2N_f)$. After chiral symmetry breaking has occurred,
$SU(2N_f)\rightarrow Sp(2N_f)$, the degrees of freedom in the
chiral sector of the effective theory are $2N_f^2-N_f-1$ Goldstone
fields $\pi^a~$, and a scalar field $\sigma$. For $N_f=2$ the
potential is \cite{Appelquist:1999dq,Lenaghan:2001sd}:
\begin{eqnarray}
V_{\rm ch}[\sigma,\pi^a]&=&\frac{m^2}{2}{\rm Tr
}\left[M^{\dagger}M\right]+ {\lambda_1}{\rm Tr
}\left[M^{\dagger}M\right]^2+ \frac{\lambda_2}{4}{\rm Tr
}\left[M^{\dagger}MM^{\dagger}M\right] \label{chiralpot}
\end{eqnarray}
with $2\,M=\sigma + i\,2\sqrt{2}\pi^a\,X^a$, $a=1,\dots,5$ and
$X^a\in {\mathcal A}(SU(4))-{\mathcal A}(Sp(4))$. The generators
$X^a$ are provided explicitly in equation (A.5) and (A.6) of
\cite{Appelquist:1999dq}. The Polyakov loop is treated as a heavy
field singlet under the chiral symmetry. Its contribution to the
potential in the absence of the $Z_2$ symmetry is
\begin{eqnarray}
V_\chi[\chi]=g_0\chi+\frac{m_\chi^2}{2}\chi^2+\frac{g_3}{3}\chi^3
+\frac{g_4}{4}\chi^4 \, . \label{chipot}
\end{eqnarray}
The field $\chi$ represents the Polyakov loop itself. To complete
the effective theory we introduce interaction terms allowed by the
chiral symmetry
\begin{eqnarray}
V_{\rm{int}}[\chi,\sigma,\pi^a]&=& \left(g_1\chi
+g_2\chi^2\right){\rm Tr } \left[M^{\dagger}M\right]=\left(g_1\chi
+g_2\chi^2\right)(\sigma^2+\pi^a\pi^a) \ .\nonumber \\
\end{eqnarray}
The $g_1$ term plays a fundamental role and it is exactly the one
we already mentioned that have been dropped in previous
investigations \cite{Dumitru:2000in,{Scavenius:2002ru}}. In the
phase with $T<T_{c\sigma}$ ($T_{c\sigma}\equiv T_{\rm chiral}$),
where chiral symmetry is spontaneously broken, $\sigma$ acquires a
non-zero expectation value, which in turn induces a modification
also for $\langle\chi\rangle$. The usual choice for alignement is
in the $\sigma$ direction, i.e. $\langle\pi\rangle=0~$. The
extremum of the linearized potential is at
\begin{eqnarray}
\langle\sigma\rangle^2 &\simeq& -\frac{m^2_{\sigma}}{\lambda}\, ,
\qquad \quad\,\,\, \quad m^2_{\sigma} \simeq m^2 + 2g_1\langle
\chi\rangle \label{vevsigma} , \\ \langle \chi \rangle\,\,
&\simeq& \chi_0 -\frac{g_1}{m_\chi^2}\langle\sigma\rangle^2\, ,
\quad ~ \chi_0 \simeq -\frac{g_0}{m^2_{\chi}} \ ,
 \label{vevchi}
\end{eqnarray}
where $\lambda=\lambda_1 + \lambda_2$. The previous formulae hold
near the phase transition where $\sigma$ is small. Here
$m^2_{\sigma}$ is the full coefficient of the $\sigma^2$ term in
the tree-level Lagrangian which, due to the coupling between
$\chi$ and $\sigma$, also depends on $\langle\chi\rangle$.
Spontaneous chiral symmetry breaking appears for $m^2_{\sigma}
<0$. In this regime the positive mass squared of the $\sigma$ is
$M^2_{\sigma} = 2 \lambda\langle \sigma ^2 \rangle$. Near the
critical temperature the mass of the order parameter field is
assumed to posses the generic behavior $m_\sigma^2\sim
(T-T_{\rm{c}})^\nu$. Equation (\ref{vevchi}) shows that for
$g_1>0$ and $g_0<0$ the expectation value of $\chi$ behaves
oppositely to that of $\sigma~$: As the chiral condensate starts
to decrease towards chiral symmetry restoration, the expectation
value of the Polyakov loop starts to increase, signaling the onset
of deconfinement. This is illustrated in the left panel of figure
\ref{Figura1}. Positivity of the expectation values implies
$2g_1^2-\lambda m_\chi^2<0$, which also makes the extremum a
minimum. At the one-loop level one can show \cite{Mocsy:2003un}
that also $\chi_0$ acquires a temperature dependence. When
applying the analysis presented in
\cite{Mocsy:2003tr,{Mocsy:2003un}}, the general behavior of the
spatial two-point correlator of the Polyakov loop can be
determined. Near the transition point, in the broken phase, the
$\chi$ two-point function is dominated by the infrared divergent
$\sigma$-loop. This is so, because the $\pi^a$ Goldstone fields
couple only derivatively to $\chi$, and thus decouple. We find a
drop in the screening mass of the Polyakov loop at the phase
transition. When approaching the transition from the unbroken
phase the Goldstone fields do not decouple, but follow the
$\sigma$, resulting again in the drop of the screening mass of the
Polyakov loop (this is actually the string tension) close to the
phase transition. Consider the variation of the $\chi$ mass near
the phase transition with respect to the tree level mass,
$m_{\chi}~$, defined above the chiral phase transition, $\Delta
m_\chi^2(T)=m_\chi^2(T)-m_\chi^2~$. Using a large $N$ framework
motivated resummation \cite{Mocsy:2003un} we deduce:
\begin{eqnarray}
\Delta m^2_\chi(T)&=& -
\frac{2g_1^2(1+N_\pi)}{8\pi\,m_{\sigma}+(1+N_\pi)3\lambda }
 \, , \quad T>T_{\rm{c}\sigma} \\
\Delta m^2_\chi(T)&=& - \frac{2g_1^2}{8\pi\,M_{\sigma}+ 3\lambda}
 \, , \qquad \qquad \,\,\,\, T<T_{\rm{c}\sigma}\, .
\end{eqnarray}
{}From the above equations one finds that the screening mass of the
Polyakov loop is continuous and finite at $T_{\rm{c}\sigma}$, and
$\Delta m_\chi^2(T_{\rm{c}\sigma})=-2g_1^2/(3\lambda)$,
independent of $N_\pi$, the number of pions. This analysis is not
restricted to the chiral/deconfining phase transition. The
entanglement between the order parameter (the chiral condensate)
and the non-order parameter field (the Polyakov loop) is
universal.

%%%%%%%%%%%%%%%%%%%%%%%%%%%%%%%%%%%
\subsection{Adjoint Representation}
As a second application, consider two color QCD with two massless
Dirac quark flavors in the adjoint representation. Here the global
symmetry is $SU(2N_f)$ which breaks via a bilinear quark
condensate to $O(2N_f)$. The number of Goldstone bosons is
$2N_f^2+N_f-1$. We take $N_f=2$. There are two exact order
parameter fields: the chiral $\sigma$ field and the Polyakov loop
$\chi~$. Since the relevant interaction term $g_1\chi\sigma^2$ is
now forbidden, one might expect no efficient information transfer
between the fields. While respecting general expectations the
following analysis suggests the presence of a new and more
elaborated structure which lattice data can clarify in the near
future. The chiral part of the potential is given by
(\ref{chiralpot}) with $2\,M=\sigma + i\,2\sqrt{2}\pi^a\,X^a$,
$a=1,\dots,9$ and $X^a\in {\mathcal A}(SU(4))-{\mathcal A}(O(4))$.
$X^a$ are the generators provided explicitly in equation (A.3) and
(A.5) of \cite{Appelquist:1999dq}. The now $Z_2$ symmetric
potential for the Polyakov loop is
\begin{eqnarray}
V_\chi[\chi]=\frac{m_{0\chi}^2}{2}\chi^2+\frac{g_4}{4}\chi^4 \, ,
\end{eqnarray}
and the only interaction term allowed by symmetries is
\begin{equation}
V_{\rm{int}}[\chi,\sigma,\pi]=g_2\chi^2\,{\rm Tr
}\left[M^{\dagger}M\right]
 =g_2\chi^2(\sigma^2+\pi^a\pi^a) \, .
\end{equation}
\begin{figure}[t]
\includegraphics[width=11truecm, clip=true]{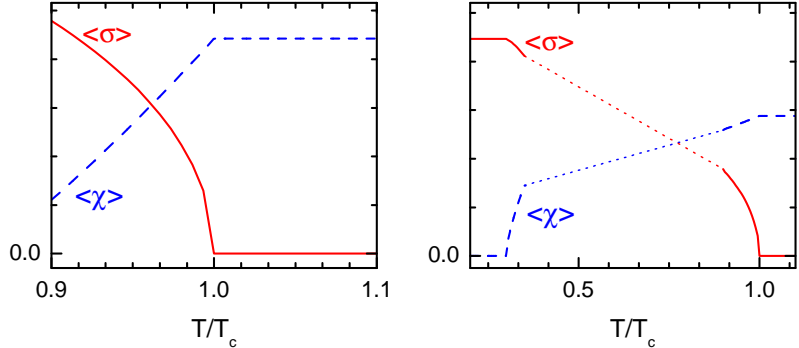}
\caption{Left panel: Behavior of the expectation values of the
Polyakov loop and chiral condensate close to the chiral phase
transition as a function of temperature, with massless quarks in
the fundamental representation. Right panel: Same as in left
panel, for massless quarks in the adjoint representation and
$T_{\rm{c}\chi}\ll T_{\rm{c}\sigma}$ (see discussion in the
text).} \label{Figura1}
\end{figure}
The effective Lagrangian does not know which transition happens
first but this is irrelevant for the validity of our general
results. Consider the physical case in which the deconfinement
phase transition happens first. Denoting the deconfining critical
temperature with $T_{\rm{c}\chi}$ and the chiral critical
temperature with $T_{\rm{c}\sigma}~$, we put ourselves in the
limit $T_{{\rm{c}}\chi}\ll T_{{\rm{c}}\sigma}$. {}For
$T_{{\rm{c}}\chi}<T<T_{{\rm{c}}\sigma}$ both symmetries are
broken, and the expectation values of the two order parameter
fields are linked to each other:
\begin{eqnarray}
\langle\sigma\rangle^2&=&-\frac{1}{\lambda}\left(m^2+
2g_2\langle\chi\rangle^2\right)\equiv
-\frac{m_\sigma^2}{\lambda}\, ,\nonumber
\\
\langle\chi\rangle^2&=&-\frac{1}{g_4}\left(m_{0\chi}^2+
2g_2\langle\sigma\rangle^2\right)\equiv -\frac{m_\chi^2}{g_4}\, .
\label{vevad}
\end{eqnarray}
The coupling $g_2$ is taken to be positive. The expected behavior
of $m_\chi^2\sim (T-T_{\rm{c}\chi})^{\nu_{\chi}}$ and
$m_\sigma^2\sim (T-T_{\rm{c}\sigma})^{\nu_{\sigma}}$ near
$T_{\rm{c}\chi}$ and $T_{\rm{c}\sigma}$, respectively, combined
with the result of eq. (\ref{vevad}), yields in the neighborhood
of these two transitions the qualitative situation illustrated in
the right panel of figure \ref{Figura1}. On both sides of
$T_{\rm{c}\chi}$ the relevant interaction term $g_2\langle
\sigma\rangle\sigma\chi^2$ emerges, leading to a one-loop
contribution to the static two-point function of the $\sigma$
field $\propto \langle \sigma \rangle^2 /m_\chi~$. Near the
deconfinement transition $m_\chi\rightarrow 0$ yielding an
infrared sensitive screening mass for $\sigma$. Similarly, on both
sides of $T_{\rm{c}\sigma}$ the interaction term $\langle
\chi\rangle\chi\sigma^2$ is generated, leading to the infrared
sensitive contribution $\propto \langle \chi\rangle^2/m_\sigma$ to
the $\chi$ two-point function. We conclude, that when
$T_{\rm{c}\chi}\ll T_{\rm{c}\sigma}$, the two order parameter
fields, a priori unrelated, do feel each other near the respective
phase transitions. It is important to emphasize that the effective
theory works only in the vicinity of the two phase transitions.
Interpolation through the intermediate temperature range is shown
by dotted lines in the right panel of figure \ref{Figura1}.
Possible structures here must be determined via first principle
lattice calculations. The infrared sensitivity leads to a drop in
the screening masses of each field in the neighborhood of the
transition of the other, which becomes critical, namely of the
$\sigma$ field close to $T_{\rm{c}\chi}$, and of the $\chi$ field
close to $T_{\rm{c}\sigma}~$. The resummation procedure outlined
in the previous section predicts again a finite drop for the
masses:
\begin{eqnarray}
\Delta m^2_\chi(T_{\rm{c}\sigma})=-\frac{8g_2^2\langle\chi\rangle
^2}{3\lambda}, \quad \Delta
m^2_{\sigma}(T_{\rm{c}\chi})=-\frac{8g_2^2\langle\sigma\rangle
^2}{3g_4}\, .
\end{eqnarray}
We thus predict the existence of substructures near these
transitions, when considering fermions in the adjoint
representation. Searching for such hidden behaviors in lattice
simulations would help to further understand the nature of phase
transitions in QCD.

%%%%%%%%%%%%%%%%%%%%%%%%%%%%%%%%%%%%%
\subsection{Quark Chemical Potential}
The analysis can be extended for phase transitions driven by a
chemical potential. In fact, for two color QCD this is
straightforward to show. When considering fermions in the
pseudoreal representation there is a phase transition from a
quark-antiquark condensate to a diquark condensate
\cite{Hands:2001jn}. We hence predict, in two color QCD, that when
diquarks form for $\mu=m_{\pi}$, the Polyakov loop also feels the
presence of the phase transition exactly in the same manner as it
feels when considering the temperature driven phase transition.
Such a situation is supported by recent lattice simulations
\cite{Alles:2002st}.

%%%%%%%%%%%%%%%%%%%%%%%%%%%%%%%%%
\section{Conclusions and Outlook}
We have shown how deconfinement (i.e. a rise in the Polyakov loop)
is a consequence of chiral symmetry restoration in the presence of
massless fermions in the fundamental presentation. In nature
quarks have small, but non-zero masses, which makes chiral
symmetry only approximate. Nevertheless, the picture presented
here still holds: confinement is driven by the dynamics of the
chiral transition in the chiral limit. The argument can be
extended even further: If quark masses were very large then chiral
symmetry would be badly broken, and could not be used to
characterize the phase transition. But in such a case the $Z_2$
symmetry becomes more exact, and by reversing the roles of the
protagonists in the previous discussion, we would find that the
$Z_2$ breaking drives the (approximate) restoration of chiral
symmetry. Which of the underlying symmetries demands and which
amends can be determined directly from the critical behavior of
the spatial correlators of hadrons or of the Polyakov loop
\cite{Mocsy:2003tr,{Mocsy:2003un}}. With quarks in the adjoint
representation we investigated the physical scenario
\cite{Karsch:1998qj} in which chiral symmetry is restored after
deconfinement sets in. In this case we have pointed to the
existence of an interesting structure, which was hidden until now:
There are still two distinct phase transitions, but since the
fields are now entangled, the transitions are not independent.
This entanglement is shown at the level of expectation values and
spatial correlators of the fields. More specifically, the spatial
correlator of the field which is not at its critical temperature
will in any case feel the phase transition measured by the other
field. Lattice simulations will play an important role in checking
these predictions. The results presented here are not limited to
describing the chiral/deconfining phase transition and can readily
be used to understand phase transitions sharing similar features.

The effective Lagrangians shown here can be immediately used,
following \cite{Dumitru:2000in,{Scavenius:2002ru}}, to study the
physics at RHIC. We expect the new trilinear term
\cite{Mocsy:2003un}, essential for our understanding of the
relation between confinement and chiral symmetry breaking
\cite{Mocsy:2003qw}, to play a role also for the physics of heavy
ion collisions.

On a general level we have shown that the critical behavior
encoded in the order parameter field is carried over the non-order
parameter field. By monitoring spatial correlators not associated
to the order parameter field one can determine the critical
behavior of the theory.

\vskip.2cm \centerline{\bf Acknowledgments}

We thank P. Damgaard, A. Jackson, J. Kapusta, J. Schechter and B.
Tomasik for discussions and careful reading of the manuscript. We
thank P. Petreczky and R. Pisarski for insightful discussions.
F.S. thanks the organizers for having arranged such an interesting
and stimulating conference.

\end{document}